\documentclass[letter,traditabstract,longauth,letterpaper]{aa}  

\usepackage{graphicx}
\usepackage{txfonts}
\usepackage{natbib}
\begin{document}
   \title{The Herschel Lensing Survey (HLS):
     Overview\thanks{{\it Herschel} is an ESA space observatory with science
       instruments provided by European-led Principal Investigator
       consortia and with important participation from
       NASA.  Data presented in this paper were analyzed using
       ``The {\it Herschel} Interactive Processing Environment (HIPE),'' a
       joint development by the {\it Herschel} Science Ground Segment
       Consortium, consisting of ESA, the NASA {\it Herschel} Science
       Center, and the HIFI, PACS and SPIRE consortia.}}


   \author{
          E.~Egami\inst{\ref{inst1}}
          \and
          M.~Rex\inst{\ref{inst1}}
          \and
          T.~D.~Rawle\inst{\ref{inst1}}
          \and
          P.~G.~P\'{e}rez-Gonz\'{a}lez\inst{\ref{inst14},\ref{inst1}}
          \and
          J.~Richard\inst{\ref{inst15}}
          \and
          J.-P.~Kneib\inst{\ref{inst6}}
          \and
          D.~Schaerer\inst{\ref{inst8},\ref{inst5}}
          \and
          B.~Altieri\inst{\ref{inst2}}
          \and
          I.~Valtchanov\inst{\ref{inst2}}
          \and
          A.~W.~Blain\inst{\ref{inst3}}
          \and
          D.~Fadda\inst{\ref{inst9}}
          \and
          M.~Zemcov\inst{\ref{inst3},\ref{inst4}}
          \and
          J.~J.~Bock\inst{\ref{inst3},\ref{inst4}}
          \and
          F.~Boone\inst{\ref{inst5},\ref{inst7}}
          \and
          C.~R.~Bridge\inst{\ref{inst3}}
          \and
          B.~Clement\inst{\ref{inst6}}
          \and
          F.~Combes\inst{\ref{inst7}}
          \and
          C.~D.~Dowell\inst{\ref{inst3},\ref{inst4}}
          \and
          M.~Dessauges-Zavadsky\inst{\ref{inst8}} 
          \and
          O.~Ilbert\inst{\ref{inst6}}
          \and
          R.~J.~Ivison\inst{\ref{inst10},\ref{inst11}}
          \and
          M.~Jauzac\inst{\ref{inst6}}
          \and
          D.~Lutz\inst{\ref{inst12}}
          \and
          L.~Metcalfe\inst{\ref{inst2}}
          \and
          A.~Omont\inst{\ref{inst13}}
          \and
          R.~Pell\'{o}\inst{\ref{inst5}}
          \and
          M.~J.~Pereira\inst{\ref{inst1}}
          \and
          G.~H.~Rieke\inst{\ref{inst1}}
          \and
          G.~Rodighiero\inst{\ref{inst16}}
          \and
          I.~Smail\inst{\ref{inst15}}
          \and
          G.~P.~Smith\inst{\ref{inst17}}
          \and
          G.~Tramoy\inst{\ref{inst6}}
          \and
          G.~L.~Walth\inst{\ref{inst1}}
          \and
          P.~van~der~Werf\inst{\ref{inst18}}
          \and
          M.~W.~Werner\inst{\ref{inst4}}
          }

   \institute{Steward Observatory, University of Arizona,
              933 N. Cherry Ave, Tucson, AZ 85721, USA;
              \email{eegami@as.arizona.edu}\label{inst1}
         \and
         Departamento de Astrof\'{\i}sica, Facultad de
         CC. F\'{\i}sicas, Universidad Complutense de Madrid, E-28040
         Madrid, Spain\label{inst14}
         \and
         Institute for Computational Cosmology, Department of Physics,
         Durham University, South Road, Durham DH1 3LE, UK\label{inst15}
         \and
         Laboratoire d'Astrophysique de Marseille, CNRS -
         Universit\'{e} Aix-Marseille, 38 rue Fr\'{e}d\'{e}ric
         Joliot-Curie, 13388 Marseille Cedex 13, France\label{inst6}
         \and
         Geneva Observatory, University of Geneva, 51, Ch. des
         Maillettes, CH-1290 Versoix, Switzerland\label{inst8}
         \and
         Laboratoire d'Astrophysique de Toulouse-Tarbes,
         Universit\'{e} de Toulouse, CNRS, 14 Av. Edouard Belin, 31400
         Toulouse, France\label{inst5}
         \and
         {\it Herschel} Science Centre, ESAC, ESA, PO Box 78, Villanueva de
         la Ca\~nada, 28691 Madrid, Spain\label{inst2}
         \and
         California Institute of Technology, Pasadena, CA 91125,
         USA\label{inst3}
         \and
         NASA {\it Herschel} Science Center, California Institute of
         Technology, MS 100-22, Pasadena, CA 91125, USA\label{inst9}
         \and
         Jet Propulsion Laboratory, Pasadena, CA 91109, USA\label{inst4}
         \and
         Observatoire de Paris, LERMA, 61 Av. de l'Observatoire, 75014
         Paris, France\label{inst7}
         \and
         UK Astronomy Technology Centre, Science and Technology
         Facilities Council, Royal Observatory, Blackford Hill,
         Edinburgh EH9 3HJ, UK\label{inst10}
         \and
         Institute for Astronomy, University of Edinburgh, Blackford
         Hill, Edinburgh EH9 3HJ, UK\label{inst11}
         \and
         Max-Planck-Institut f\"{u}r extraterrestrische Physik,
         Postfach 1312, 85741 Garching, Germany\label{inst12}
         \and
         Institut d'Astrophysique de Paris, CNRS and Universit\'{e}
         Pierre et Marie Curie, 98bis Boulevard Arago, F-75014 Paris,
         France\label{inst13}
         \and
         Department of Astronomy, University of Padova,
         Vicolo dell'Osservatorio 3, I-35122 Padova, Italy\label{inst16}
         \and
         School of Physics and Astronomy, University of Birmingham,
         Edgbaston, Birmingham, B15 2TT, UK\label{inst17}
         \and
         Sterrewacht Leiden, Leiden University, PO Box 9513, 2300 RA
         Leiden, the Netherlands\label{inst18}
         }

   \date{Received ; accepted }

 
  \abstract{The {\it Herschel} Lensing Survey (HLS) will conduct deep
    PACS and SPIRE imaging of $\sim$40 massive clusters of galaxies.
    The strong gravitational lensing power of these clusters will
    enable us to penetrate through the confusion noise, which sets the
    ultimate limit on our ability to probe the Universe with {\it
      Herschel}.  Here, we present an overview of our survey and a
    summary of the major results from our Science Demonstration Phase
    (SDP) observations of the Bullet Cluster ($z=0.297$).  The SDP
    data are rich, allowing us to study not only the background
    high-redshift galaxies (e.g., strongly lensed and distorted
    galaxies at $z=$2.8 and 3.2) but also the properties of
    cluster-member galaxies.  Our preliminary analysis shows a great
    diversity of far-infrared/submillimeter spectral energy
    distributions (SEDs), indicating that we have much to learn with
    {\it Herschel} about the properties of galaxy SEDs.  We have also
    detected the Sunyaev-Zel'dovich (SZ) effect increment with the
    SPIRE data.  The success of this SDP program demonstrates the
    great potential of the {\it Herschel} Lensing Survey to produce
    exciting results in a variety of science areas.}


   \maketitle


\section{Introduction}

With the successful launch and commissioning of the ESA's {\it
  Herschel} Space Observatory \citep{Pilbratt10}, we are again on the
verge of making great new discoveries.  Following the
  breakthrough submillimeter/millimeter observations with SCUBA and
  MAMBO \citep[c.f.,][]{Blain02}, deep MIPS 24 $\mu$m observations
carried out by the {\it Spitzer} Space Telescope have enabled us to
trace the evolution of infrared-luminous galaxies up to $z \sim$ 3--4
\citep[e.g.,][]{Pgperez05,LeFloch05}.  However, the validity of all
these {\it Spitzer}-based results rests on the assumption that the
total infrared luminosities of high-redshift infrared galaxies can be
estimated accurately by sampling their rest-frame mid-infrared
emission (e.g., the MIPS 24$\mu$m band samples the rest-frame 8 $\mu$m
emission at $z=2$).  Indeed, some {\it Spitzer} results have already
questioned this assumption, suggesting that the use of local galaxy
spectral energy distribution (SED) templates may lead to
overestimating the total infrared luminosities of high-redshift
galaxies \citep[e.g.,][]{Papovich07,Rigby08}.  {\it Herschel} will
allow us to measure the total infrared luminosities of a large number
of high-redshift galaxies directly for the first time.

With {\it Herschel}, confusion noise produced by a sea of blended faint
galaxies sets the ultimate limit on how deeply we can probe the
Universe.  Once the source confusion sets in, it is no longer possible
to improve the detection limit by integrating longer.  This limitation
is especially severe for SPIRE, which reaches the confusion limit
quickly.

To penetrate through the confusion limit, gravitational lensing by
massive galaxy clusters offers a very powerful and yet cheap solution
\citep[e.g.,][]{Blain97}.  Magnification factors of 2--4x are quite
common in the cluster core regions, and when a background source is
strongly lensed (i.e., multiply imaged), magnification factors can
reach 10x--30x or more.  Note that a magnification factor of 10x
corresponds to a factor of 100x saving in observing time when the
sensitivity is background-limited.  Therefore, a fairly
short-integration image of a cluster core region would often reveal
sources that are well below the detection limit of an ultra-deep
blank-field image.  This method was for example employed for the first
SCUBA observations of the high-redshift Universe, which resulted in
the identification of the substantial population of infrared-luminous
galaxies at $z>1$ \citep{Smail97}.

The use of gravitational lensing is especially powerful at
infrared/submillimeter wavelengths. This is because cluster cores are
dominated by early-type galaxies, which usually emit little at these
wavelengths.  Therefore, infrared/submillimeter sources detected in
cluster cores are often background galaxies.  In other words, when
observed at these wavelengths, massive cluster cores virtually act as
a transparent lens.  Lensing studies in the infrared/submillimeter
also benefits from the steep galaxy counts at these wavelengths.

These types of lensing surveys, however, have one limitation: the
small number of strongly lensed galaxies observed per cluster.
Although a large number of massive clusters have been targeted by
ground-based submillimeter/millimeter observations
\citep[e.g.,][]{Smail02,Chapman02,Knudsen08}, the number of strongly
lensed (i.e., multiply imaged) galaxies discovered in these surveys
remains small: for example, the $z=2.5$ galaxy in Abell 2218
\citep{Kneib04}, the $z=2.9$ galaxy in MS0451.6-0305 \citep{Borys04},
the $z=2.8$ galaxy in the Bullet Cluster
\citep{Wilson08,Gonzalez09,Rex09,Johansson10}, and most recently the
exceptionally bright $z=2.32$ galaxy in MACSJ2135-010217
\citep{Swinbank10}.  Based on these observations, we empirically
  estimate the rate of finding such strongly lensed
  infrared/submillimeter galaxies to be roughly 1 in 10 with the
  sensitivity of ground submillimeter observations (e.g., SCUBA,
  LABOCA).  Therefore, many tens of clusters need to be observed to
  study a significant number of strongly lensed galaxies.

\section{The Herschel Lensing Survey (HLS)}

As a {\it Herschel} Open-Time Key Program, we are conducting exactly
such a large lensing survey, targeting $\sim$40 massive clusters of
galaxies (``The Herschel Lensing Survey (HLS)'', PI - Egami, 272.3
hrs).  Together with the PACS and SPIRE Guaranteed-Time teams, which
will observe 10 clusters \citep{Altieri10,Blain10}, we will obtain
deep images with PACS \citep{Poglitsch10} at 100 and 160 $\mu$m and
SPIRE \citep{Griffin10} at 250, 350, and 500 $\mu$m for a sample of
$\sim$50 massive clusters as a legacy of the {\it Herschel} mission.


{\bf Target Selection} --- As the targets of the survey, we have
chosen the most X-ray-luminous clusters from the ROSAT X-ray all-sky
survey assuming that the most X-ray-luminous clusters are also the
most massive and therefore the most effective gravitational lenses.
The majority of our targets come from the sample of the Local Cluster
Substructure Survey (LoCuSS) \citep[e.g.,][]{Smith10}, which adopts
the following selection criteria: (1) $L_{X} > 2\times10^{44}$ erg/s,
(2) $0.15<z<0.3$, (3) $N_{HI} < 7\times10^{20}$ cm$^{-2}$, and (4)
$-70<\delta<70$.  In addition, some number of clusters with
spectacular lensed systems were included in the sample.  For the
majority of our target clusters, we have well-constrained accurate
mass models, which have been constructed through many years of
intensive imaging/spectroscopic campaigns with HST, Keck, and VLT
telescopes.  Other important considerations were the availability of
MIPS 24 $\mu$m images and accessibility from ALMA for future follow-up
observations although these conditions were not always met.


{\bf Observing Parameters} --- Each target cluster is imaged by both
PACS (100 and 160 $\mu$m) and SPIRE (250, 350, and 500 $\mu$m).  With
PACS, we use the scan-map mode with the medium speed (some early data
were taken with the slow speed).  The scan leg lengths are 4\arcmin,
cross-scan step is 20\arcsec, number of scan legs is 13.  Each cluster
is observed twice by orthogonal scan maps (map orientation angles of
45\degr\ and 315\degr) with 18 repetitions each.  The total observing
time is 4.4 hrs per cluster with an on-source integration time of 1.6
hrs (1500 sec/pixel).

With SPIRE, we use the Large Map mode with the nominal speed.  The
scan direction was set to Scan Angles A and B.  The length and height
of the map are set to 4\arcmin, which in practice will produce a map
of 17\arcmin$\times$17\arcmin.  With 20 repetitions, the total
observing time per cluster is 1.7 hrs with an on-source integration
time of 0.6 hrs (17 sec/pixel).

{\bf Coordinated Programs} --- The {\it Herschel} Lensing Survey is
directly coordinated with a few other observing programs.  The most
important is the {\it Spitzer}/IRAC Lensing Survey (PID 60034; ``The
IRAC Lensing Survey: Achieving JWST Depth with Spitzer'', PI - Egami,
526 hrs), which is one of the {\it Spitzer} Warm-Mission
Exploration Science programs.  This program will obtain deep
(5hr/band) {\it Spitzer}/IRAC 3.6 and 4.5 $\mu$m images of $\sim$50
massive clusters.  By design, its target list is highly overlapped
with that of the HLS.  Deep IRAC images will be essential for
identifying optically-faint high-redshift infrared-luminous galaxies
as well as for deriving accurate photometric redshifts.  In addition,
roughly half of the HLS clusters are being imaged by HST/WFC3 through
two on-going programs (GO 11592: ``Are Low-Luminosity Galaxeis
Responsible for Reionization?'', PI - Kneib, 43 orbits; MCT: ``Through
a Lens, Darkly - New Constraints on the Fundamental Components of the
Cosmos'', PI - Postman, 524 orbits).

The {\it Herschel} Lensing Survey is also closely related to two other
{\it Herschel} Open-Time Key Programs: ``LoCuSS: A Legacy Survey of
Galaxy Clusters at $z=0.2$'' \citep{Smith10} and ``Constraining the
Cold Gas and Dust in Cluster Cooling Flows'' \citep{Edge10}.  The
former will obtain wide (30\arcmin$\times$30\arcmin) and shallow PACS
100/160 $\mu$m maps of $\sim$30 massive galaxy clusters at $z \sim
0.2$, many of which are also targeted by the HLS.  Note that the HLS
SPIRE maps cover a significant part (the central
17\arcmin$\times$17\arcmin) of the LoCuSS PACS maps, leading to a
natural collaboration between the two teams.  The latter program will
study the brightest cluster galaxies (BCGs) in a dozen cooling-flow
clusters, and the HLS data will provide PACS/SPIRE photometry for a
much larger sample of BCGs.

\section{Science Demonstration Target: The Bullet Cluster}


\begin{figure*}[bth]
     \includegraphics[width=4.5in]{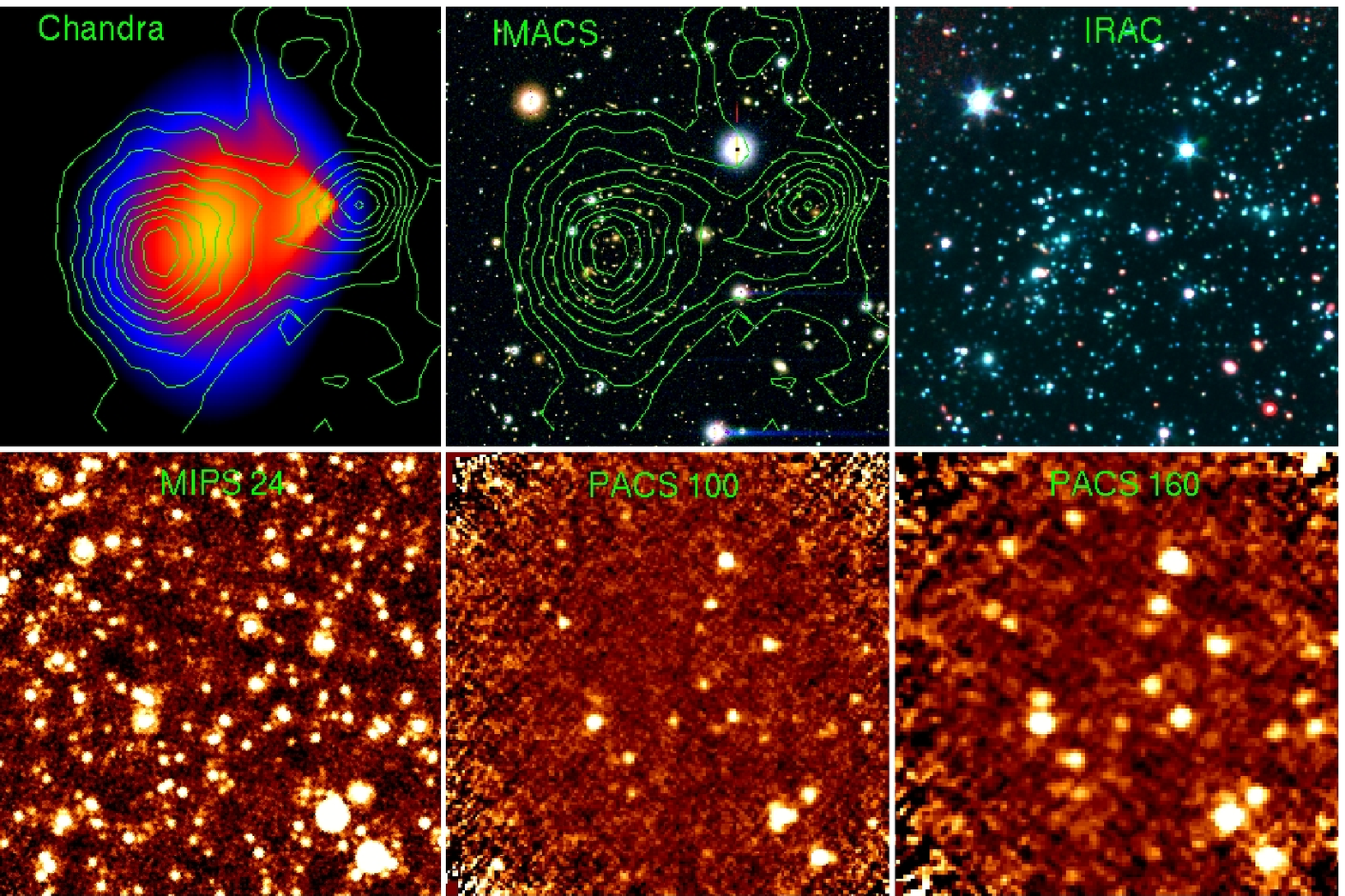}

     \vspace*{-6.55cm}

    \includegraphics[width=4.5in]{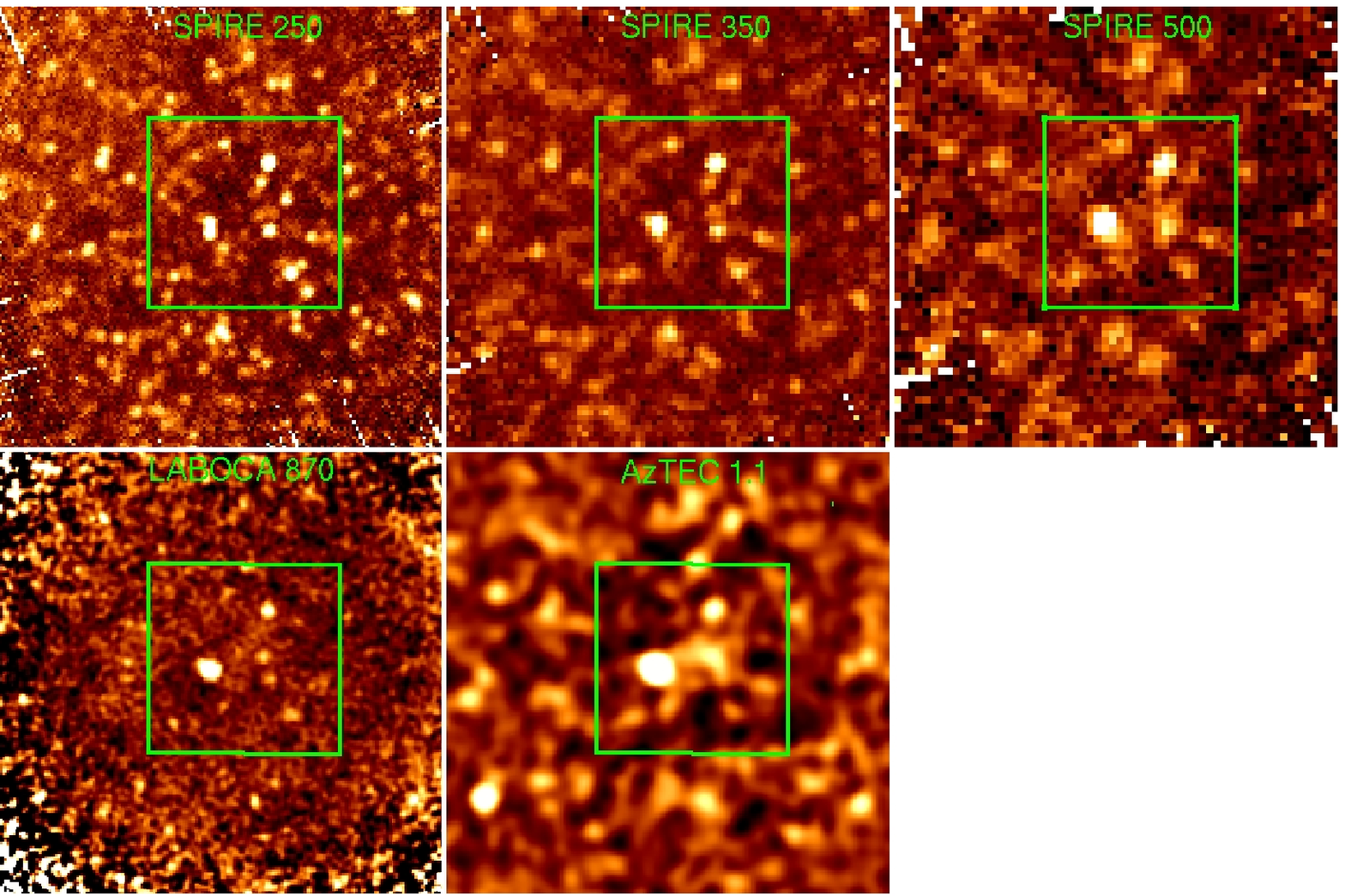}
    \includegraphics[width=2.6in]{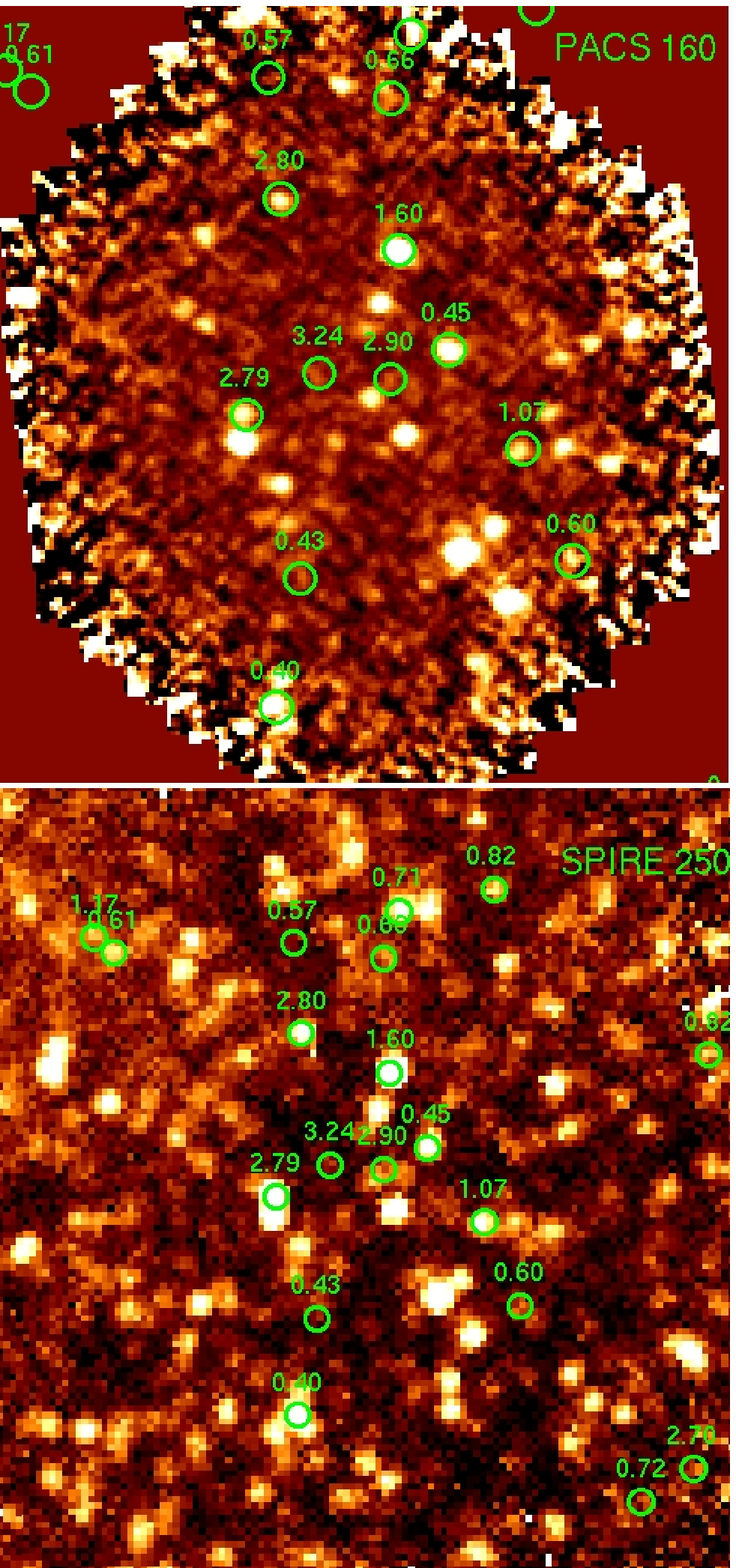}

    \caption{Left -- Multi-wavelength imaging data of the Bullet
      Cluster as indicated in each panel (see Appendix A for
      references).  The contours overlaid on the Chandra and IMACS
      images are the weak-lensing mass map from \citet{Clowe06}.  The
      field of view is $\sim$7\arcmin$\times$7\arcmin\ in the first
      two rows and $\sim$15\arcmin$\times$15\arcmin\ in the last two
      rows (the $\sim$7\arcmin$\times$7\arcmin\ area is indicated with
      the squares); Right -- Spectroscopic and photometric redshifts
      overlaid on the PACS 160 $\mu$m map (top) and SPIRE 250 $\mu$m
      map (bottom).  See \citet{Rex10} for the full source list.}

    \label{Herschel_maps}
\end{figure*}

In the Science Demonstration Phase (SDP), we observed the Bullet
Cluster at $z=0.297$ (1E0657-56$=$RXCJ~0658.5-5556).  The Bullet
Cluster was targeted because, (1) there is a strongly lensed bright
submillimeter/millimeter galaxy at $z=2.8$ (see the Introduction), (2)
it has a large amount of ancillary data (see Appendix~A), and (3) it
is in the continuous viewing zone of {\it Herschel}.

\subsection{Data Processing}

\begin{table}[b]
\caption{The HLS Bullet Cluster Data}
\label{table:data}
\centering
\begin{tabular}{lcccc} \hline\hline
Instrument &  FOV              & $\lambda$ & beam     &  Depth     \\ 
           & (arcmin)          & ($\mu$m)  & (arcsec) & (5$\sigma$, mJy)       \\ \hline
PACS       &  $\sim8\times8$   &   100     & 7.7      & 5.5$\pm$0.7  \\ 
           &                   &   160     & 12.0     &  10$\pm$1    \\
SPIRE      &  $\sim17\times17$ &   250     & 18       &  12$\pm$2  \\ 
           &                   &   350     & 25       &  17$\pm$3  \\ 
           &                   &   500     & 36       &  18$\pm$4  \\  \hline
\end{tabular}
\end{table}

Raw {\it Herschel} data products are reduced using the common pipeline
procedures distributed within the {\it Herschel} Interactive Processing
Environment (HIPE) \citep{Ott10}. Deviations from the standard routines
are described below.

{\bf PACS} --- The PACS observations were affected by erroneous
flashes of the calibration lamp at the end of each scan
repetition. These high intensity spikes in the detector time-streams
were followed by an exponential decay in the signal while the
bolometers re-thermalized. While the spike and a large fraction of the
decay occurred during the slew between repetitions, the bolometers
were not fully stabilized until after the beginning of the next scan,
resulting in a $\sim$10 \% reduction of usable data. 

The {\it 1/f} noise drift in the PACS bolometers was removed by a high-pass
filter. Before applying the filter, bright sources ($>$ 5$\,\sigma$)
were masked to prevent Fourier ringing about their positions. These
sources were selected from an unmasked first-pass of the 160 $\mu$m
map, and the allocated mask size was proportional to their
significance. High pass filter lengths of 30 and 40 time-stream frames
were used for PACS 100/160 $\mu$m data respectively. These lengths
were selected to minimize residual {\it 1/f} noise in the resultant maps
without clipping power on the scale of the PSF. The maps incorporate
all data observed while the telescope maintained the nominal scan
speed (including some turnaround data). The final PACS maps have pixel
sizes of 2\arcsec\ and 3\arcsec\ (100/160 $\mu$m).

{\bf SPIRE} --- In addition to the nominal scan legs (speed =
30\arcsec\ s$^{-1}$), we include all turnaround data observed while
the telescope is scanning faster than 0\farcs5 s$^{-1}$, greatly
increasing coverage of the outer regions. Low-frequency drifts in the
SPIRE detectors were removed by subtracting the median value of the
nominal scan-speed data from each individual scan leg
independently. The final SPIRE maps have pixel sizes of 6\arcsec,
9\arcsec, and 12\arcsec\ (250/350/500 $\mu$m respectively).

The properties of the processed PACS and SPIRE maps are summarized in
Table~\ref{table:data}.

\subsection{Overview of the Bullet Cluster SDP Data}

Figure~\ref{Herschel_maps} shows the PACS and SPIRE images of the
Bullet cluster.  Also shown are the Chandra, Magellan/IMACS, {\it
  Spitzer}/IRAC \& MIPS 24 $\mu$m, LABOCA and AzTEC images of the same
field (see Appendix A for references).  As expected, the MIPS 24
$\mu$m observation goes deep enough to detect the counterparts for
most of the PACS/SPIRE sources.  This means that one immediate goal of
{\it Herschel} will be to determine the far-infrared SEDs of 24
$\mu$m-detected sources.  In fact, the MIPS 24 $\mu$m image enables us
to extract fluxes from confused {\it Herschel} sources accurately
\citep{Pgperez10}.  However, the most interesting type of {\it
  Herschel} sources may be those without 24 $\mu$m counterparts (such
a source has not yet been identified in our data).


In the Bullet Cluster field, we have detected two significantly lensed
galaxies, one at $z=2.79$ and the other at $z=3.24$
\citep{Rex10,Pgperez10}.  With magnification factors of $>$54 and 11.3
\citep{Paraficz10}, respectively, their intrinsic total infrared
luminosities are $<5\times10^{11} L_{\sun}$ and 3.5$\times10^{11}
L_{\sun}$.  Figure~\ref{Herschel_maps} shows the locations of these
two and other background galaxies studied.  Note that only HLS can
detect luminous infrared galaxies (LIRGs;
$10^{11}$$<$$L_{TIR}$$<$$10^{12}$ $L_{\sun}$) at $z$$>$2--3 in both
PACS and SPIRE data.

\begin{figure}[h]
  \centering
  \includegraphics[width=3.2in]{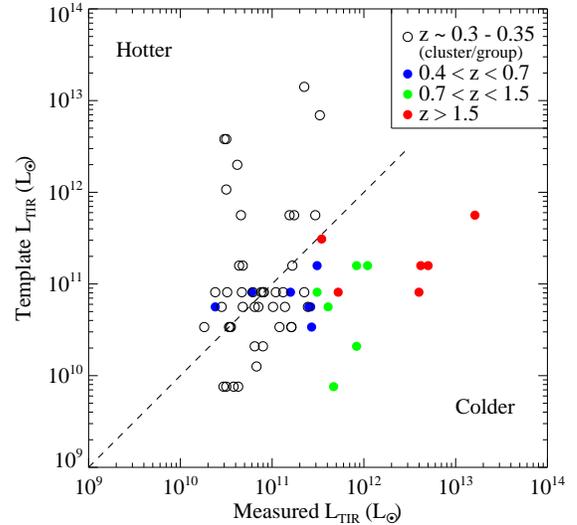}
  \caption{Comparison between the total infrared luminosities measured
    by fitting the \citet{Rieke09} SED templates to the observed data
    points (Measured $L_{TIR}$) and the total infrared luminosity
    classes originally assigned to these SED templates (Template
    $L_{TIR}$).  The open circles denote cluster/group-member galaxies
    while the solid circles denote background galaxies.  Galaxies
    below the 1:1 line have SEDs that resemble lower-luminosity local
    SED templates (i.e., colder dust temperatures, smaller
    mid-infrared to far-infrared luminosity ratios).  Galaxies above
    the line have the opposite characteristics.  The distribution of
    the points along the Y axis is discreet because of the finite
    number of templates used. See \citet{Rex10} and
    \citet{Rawle10} for more detail.}
  
  \label{lumiplot}
\end{figure}

One major result coming out of this SDP program is the diversity of
far-infrared/submillimeter SEDs seen with the $z$$=$0.3 cluster-member
galaxies, $z$$=$0.35 background group galaxies, and higher-redshift
field galaxies behind these two galaxy concentrations \citep{Rawle10,
  Rex10}.  This is shown in Figure~\ref{lumiplot}, which compares the
total infrared luminosities measured by fitting the \citet{Rieke09}
SED templates to the observed data points (Measured $L_{TIR}$) against
the total infrared luminosity classes originally assigned to these SED
templates (Template $L_{TIR}$).  Although these two luminosities agree
for many of the cluster and background group members, Template
$L_{TIR}$ is significantly lower than Measured $L_{TIR}$ for most of
the higher-redshift background galaxies.  What this means is that the
shapes of the infrared/submillimeter SEDs of these galaxies
resemble those of lower-luminosity galaxies in the local Universe.
This result is consistent with the earlier findings by various {\it
  Spitzer} observations \citep[e.g.,][]{Papovich07, Rigby08}.  This
also implies that these higher-redshift star-forming galaxies have a
larger amount of colder dust compared to the local galaxies with
similar infrared luminosities.

In the face of this SED difference, what is surprising is the recent
finding that the total infrared luminosities can be estimated fairly
well (at least up to $z\sim1.5$) if we use the luminosity-dependent
galaxy SED templates as observed locally \citep{Elbaz10}. We have
confirmed this finding with our Bullet Cluster data.  Despite this
agreement, however, we have also found that the kind of SED mistmatch
seen in Figure~\ref{lumiplot} clearly exists in the data even at $z<1$
\citep{Rex10}.  This suggests that the good match between the observed
and 24 $\mu$m-derived total infrared luminosities does not necessarily
mean that the local SED templates are making good fits to the observed
SEDs. A more detailed study of a larger sample is required to resolve
this issue.


Equally interesting is the discovery of cluster/group-member galaxies
that show large deviations in the opposite direction \citep{Rawle10}.
In other words, the SEDs of these galaxies resemble those of
higher-lumninosity galaxies in the local Universe.  These galaxies
therefore are likely to have a larger amount of hotter dust and a more
pronounced infrared SED peak compared to the local counterparts with
similar infrared luminosities.  Similar galaxies were also found in
other clusters \citep{Pereira10,Smith10}, possibly suggesting that
this type of SEDs may be specific to the cluster environment.


Finally, we also report the first detection of the Sunyaev-Zel'dovich
(SZ) effect increment at 350 and 500 $\mu$m using the SPIRE data
\citep{Zemcov10}.  The measurements will allow us to assess the
relativistic correction to the SZ effect.

\section{Conclusions}

The SDP observations of the Bullet Cluster clearly demonstrate the
great potential of the HLS in a variety of science areas.  With the
{\it Herschel} observations nearly done, HLS is expected to make a
rapid progress in the near future. One immediate interest is whether
it can find strongly lensed sources that are bright enough to perform
spectroscopy with {\it Herschel}.  Ultimately, we will construct a
definitive sample of $\sim$50 lensing clusters with a variety of
multi-wavelength data and accurate mass models.  Such a data set can
be further exploited by future facilities such as {\it ALMA}, {\it
  JWST}, {\it SPICA}, and ground 30-meter class telescopes.

\begin{acknowledgements}

We would like to thank the following people for providing various data
sets/information to us: D.~Clowe (Magellan/IMACS images), S.~M.~Chung
and A.~H.~Gonzalez (IMACS spectroscopic redshifts), J.-G.~Cuby
(VLT/HAWKI images), D.~Johansson, C.~Horellou, and the LABOCA team
(LABOCA map), and D.~Hughes, I.~Aretxaga, and the AzTEC team (AzTEC
map and far-infrared photometric redshifts).  We would also like to
thank the NASA {\it Herschel} Science Center for its excellent user
support, and the International Space Science Institute in Berne for
their support through the International team 181.  EE would like to
thank D.~Elbaz for communicating his results before publication.

This work is based in part on observations made with {\it Herschel}, a
European Space Agency Cornerstone Mission with significant
participation by NASA.  Support for this work was provided by NASA
through an award issued by JPL/Caltech.

\end{acknowledgements}

\bibliographystyle{aa}
\bibliography{ref}

\Online

\begin{appendix}

\section{Ancillary Data for the Bullet Cluster}

\begin{table}[h]
\caption{The Bullet Cluster Ancillary Data}
\label{table:data2}
\centering
\begin{tabular}{llc} \hline\hline
Facilities     & Bands                        & Ref. \\ \hline
\multicolumn{3}{c}{Imaging} \\ \hline
Magellan/IMACS & B, V, R                      & 1 \\
HST/ACS        & F606W, F775W, F850LP         & 2 \\
VLT/HAWK-I     & Y, $J$                       & 3 \\
{\it Spitzer}/IRAC   & 3.6, 4.5, 5.8, \& 8.0 $\mu$m & 2 \\
{\it Spitzer}/MIPS   & 24 $\mu$m                    & 2 \\
LABLOCA        & 870 $\mu$m                   & 4 \\
AzTEC          & 1.1 mm                       & 5 \\ \hline
\multicolumn{3}{c}{Spectroscopy} \\ \hline
Magellan/IMACS & optical (856 targets)        & 6 \\
Blanco/Hydra   & optical (202 targets)        & 7 \\
VLT FORS       & optical (14 targets)         & 8 \\ \hline
\end{tabular}
\tablebib{ (1)~\citet{Clowe06}; (2)~\citet{Gonzalez09};
  (3)~J.-G.~Cuby, priv.~comm.; (4)~\citet{Johansson10};
  (5)~\citet{Wilson08}; (6) Chung et al., in prep; (7) D.~Fadda,
    priv.~comm.; (8) J.~Richard, priv.~comm.  }
\end{table}

\end{appendix}

\end{document}